\documentclass[11pt, oneside]{article}   	
\usepackage[hmargin=1in,vmargin=1in]{geometry}           		
\usepackage{graphicx}				
								
\usepackage{subfig}	
\usepackage{color}
\usepackage[round]{natbib}
	
\usepackage{amssymb}
\usepackage{amsmath}
\usepackage{bbm}
\usepackage{amsthm}

\title{Singular Perturbation Expansion for Utility Maximization with Order-$\epsilon$ Quadratic Transaction Costs}

\author{Shiva Chandra\thanks{Department of Finance and Risk Engineering, NYU Tandon School of Engineering,Brooklyn NY \textit{sc5470@nyu.edu}}~~and Andrew Papanicolaou\thanks{Department of Finance and Risk Engineering, NYU Tandon School of Engineering, Brooklyn NY \textit{ap1345@nyu.edu}}}

\begin{document}

\maketitle

\begin{abstract}
We present an expansion for portfolio optimization in the presence of small, instantaneous, quadratic transaction costs. Specifically, the magnitude of transaction costs has a coefficient that is of the order $\epsilon$ small, which leads to the optimization problem having an asymptotically-singular Hamilton-Jacobi-Bellman equation whose solution can be expanded in powers of $\sqrt\epsilon$. In this paper we derive explicit formulae for the first two terms of this expansion. Analysis and simulation are provided to show the behavior of this approximating solution.
\textbf{AMS Subject Codes:} 91B28, 93E20. 
\end{abstract}

\section{Introduction \label{sec:intro}}

A common model for trading in block-shaped order books is the Black-Scholes model with a quadratic penalty on trades \citep[see][]{OW2013},
\begin{align}
\nonumber
dS_t &= \mu S_t dt + \sigma S_t dZ_t\\
\nonumber
\label{eq:wpnl}
dW_t&=H_tdS_t-S_t\frac{\epsilon}{2}h_t^2dt\\
dH_t&=h_tdt\ ,
\end{align}
where $S_t$ is the mid-price of the asset or stock, $W_t$ is the investor's wealth process, $Z_t$ is a standard Brownian motion, and $H_t$ is the investor's inventory or number of shares held at time $t$. The control process, $h_t$ is the rate of change of the investor's inventory. The parameter $\mu>0$ is the average rate of return of the asset and $\sigma$ is the volatility on returns. In this paper we consider the regime of $\epsilon>0$ very small (e.g. $\epsilon=.001$). Without loss of generality we have taken the risk-free rate to be zero. The interpretation of $\epsilon>0$ is that there is a small cost of executing an order that is associated with liquidity in the order book; liquidity in the order book will determine the magnitude of transaction costs incurred.

Eq. \eqref{eq:wpnl} are the variables pertinent to our investor's objective, namely, wealth, asset price and existing position in the asset. The investor's objective is to maximize the running discounted expected utility of wealth, for which the optimal value function is given as, 

\begin{equation}
\label{eq:valueFunctionIntro}
V(w,s,\xi) = \sup_{h\in\mathcal A}\mathbb E\left[\int_{t}^{\infty} e^{-\rho (u-t)}U(W_u)du \Big|W_t = w,S_t=s,H_t=\xi\right]\ .\end{equation}
where the exponential utility $U(w)=-e^{-\gamma w}/\gamma$ is the investor's concave utility function with risk aversion $\gamma >0$, $\rho>0$ is a utility discount factor, and the admissible set $\mathcal A$ is the set of almost-everywhere finite-valued adapted controls with non-negative inventory,
\begin{equation}
\label{eq:A}
\mathcal A= \left\{h:h_t\in\mathcal F_t~~,~~|h_t|<\infty~~\hbox{and}~~H_t>0 ~~\hbox{for almost everywhere $t<\infty$}\right\}\ .
\end{equation}
The control process's value $h_t$ is the change in the number of contracts held in the risky asset at each time instant. The dynamics of $H_t$ in \eqref{eq:wpnl} act as a constraint where the investor's inventory is restricted to having no diffusion component. 

The value function in \eqref{eq:valueFunctionIntro} can be interpreted as the risk preferences of an individual or institution, in possession of an abundance of wealth with the desire to have it maintained indefinitely. In fact, we will see in the sequel that this optimization has a solution that is close to the Merton problem with very similar financial intuition. If desired, an independent stopping time can be included and the optimization can be terminated at the stopping time's arrival (e.g., upon the death of an individual the wealth will be liquidated and donated to charity). The value function in \eqref{eq:valueFunctionIntro} can also be thought of as a softening of a constraint to keep wealth non-negative over an indefinite horizon, which is not too dis-similar to the drawdown constraint and infinite investment horizon considered in \cite{grossmanZhou}. 

The challenge is to find a solution for the above optimization problem, which is non-trivial because it requires us to solve a complicated Hamilton-Jacobi-Bellman (HJB) equation. We solve for an expansion that approximates this solution. We are also able to talk about the verification, existence, uniqueness and accuracy of the solution so obtained. Comparing this HJB with that of the optimal portfolio value in the Merton case, we shall attempt to understand the nature and the magnitude of the deviations that arise for $\epsilon$ small, in order to gain more understanding on how execution costs affect portfolio choices. We hypothesize that the result to our problem shall lie in the vicinity of the Merton solution and exhibit mean reversion around it. 
\subsection{Overview of Related Work}

Optimal execution with an objective to minimize volatility and linear instantaneous execution costs is done in \cite{almgren2001optimal} and gives results in the form of an efficient frontier. A similar objective is optimized in \cite{voss2016} where the optimal policy is shown to move towards a weighted average of future target positions. The relationship between mean-reversion speeds and trading strategies is presented in \cite{garleanuPedersen}, which brings forth the importance of following a target or an ``aim portfolio". Minimum-variance hedging of a call option with linear instantaneous impact is studied in \cite{rogers2010}, which presents the order-$\sqrt\epsilon$ magnitude of the hedging error --a result that is similar to the results shown in this paper and uses the model as described by \eqref{eq:wpnl}. A related reference on singular perturbation for equations of stochastic control is \cite{kosygina2017}. A general reference for small-$\epsilon$ singular perturbation theory is \cite{fouque2011multiscale}, and more recently these techniques are applied to optimal execution with linear impact and stochastic volatility in \cite{chan2015optimal}. In \cite{shreve1994} they solve the optimization problem for the consumption in infinite horizon in the presence of a proportional transaction cost when trading between a money market account and a risky asset. An interesting take on arbitrage pricing theory in the presence of transaction costs is presented in \cite{article}, which explores the extensions to the fundamental theorems of asset pricing for approximately complete markets. Optimal investment and consumption are explained in the monograph of \cite{kabanovMarkets2008} for the case of proportional transaction costs, which includes the problem of \cite{davisNorman1990}.

\subsection{Results in this Paper}
The main result presented in Section \ref{sec:two} is a power series expansion of the solution to the HJB equation with a quadratic transaction cost of order $\epsilon$, where $\epsilon$ is asymptotically small. Our expansion can be applied to any utility function $U$ such that there is a separation of variables in the solution of the HJB equation, i.e. if $V$ is the solution of the HJB then $V(w,s,\xi) = U(w)G(s,\xi)$. Specifically, for exponential utility function we write the formal expansion of $G(s,\xi)$ as,
\begin{equation}
\label{eq: PertG}
    G = G^{(0)}+\sqrt\epsilon G^{(1)} + \epsilon G^{(2)}+\dots ,
\end{equation}
for which we are able to obtain $G^{(0)}$ and $G^{(1)}$ explicitly, and we give $G^{(2)}$ explicitly up to its derivative in $\xi$. 

The expansion lies in the vicinity of and exhibits mean reversion towards the solution of the Merton problem (case $\epsilon=0$). In Section \ref{sec:verificationAndExistence} we show verification, accuracy and existence of the solution. We take a viscosity approach which enables us to talk about existence and uniqueness. We then comment on accuracy of the expansion's zero-order term for small $\epsilon$. Section \ref{sec:monteCarlo} provides simulations to show how small $\epsilon$ and the investor's risk aversion affect the optimized processes.

\section{Main Result}
\label{sec:two}
Our objective is to calculate the optimal value of wealth invested in the risky asset at each time instant during our investment horizon. We aim to achieve a maximum value of the investor's utility.
In this section we shall outline the framework for our analysis. The main objective here is to maximize the value of wealth over an infinite time horizon. Letting the asset price and wealth processes be as they are given in \eqref{eq:wpnl}, the investor's optimal value function is,

\begin{equation}
\label{eq:valueFunction}
V(w,s,\xi) = \sup_{h\in\mathcal A}\mathbb E\left[-\frac{1}{\gamma}\int_{t}^{\infty} e^{- \rho (u-t)- \gamma W_u }du \Big|W_t = w,S_t=s,H_t=\xi\right]\ ,
\end{equation}
for any $(w,s,\xi)\in\mathbb R\times\mathbb R^+\times\mathbb R^+$, with $\mathcal A$ as defined in \eqref{eq:A}. Notice that because $\epsilon>0$ we take $H_t$ as a state variable and $h_t$ as the control variable. If $\epsilon = 0$ then we are in the setting of the classical Merton problem with $H_t$ being the control variable.

\subsection{The Merton Problem (Case $\epsilon =0$)}
We briefly review the Merton problem, which in the context of this paper is the idealistic case of zero execution costs. As there is no loss due to the costs involved in buying or selling securities, the investor can continuously maintain the optimal number of risky assets, by re-balancing his/her portfolio at each instant. Hence, no matter how far the investor is in terms of the optimal number of contracts held in the portfolio, the investor can move his/her allocation to the optimal at each time instant in a continuous manner without incurring any costs. 

The wealth process in this case is,
\begin{equation}
\label{eq:Wealth0}
    dW_t = H_tS_t \frac{dS_t}{S_t} = \tilde H_t\left(\mu dt+\sigma dZ_t\right)\ ,
\end{equation}
where $\tilde H$ is the amount of wealth in the risky asset at time $t$, and can be taken as the control variable. The investor's optimal is then,
\begin{equation}
\label{eq:valueFunction0}
V(w,s) = \sup_{\tilde H\geq 0}\mathbb E\left[-\frac{1}{\gamma}\int_{t}^{\infty} e^{- \rho (u-t)- \gamma W_u }du \Big|W_t = w,S_t=s\right]\ ,\end{equation}
where the supremum is taken over all $(\tilde H_t)_{t\geq 0}$ such that $\tilde H_t$ is non-negative, $\mathcal F_t$-adapted, and finite for all $t<\infty$. Taking a dynamic programming approach, we formulate the HJB equation,
\begin{align}
\label{eq:HJBmerton}
-\frac{1}{\gamma} e^{-\gamma w} - \rho V +\max_{\tilde H}\left(\frac{\tilde H^2\sigma^2}{2}V_{ww}+\tilde H\mu V_w\right)&=0\ .
\end{align}
We employ the ansatz $V(w) = -(c/\gamma) e^{-\gamma w}$. Using first-order conditions we find the optimal amount of wealth in the risky asset at time $t$,
\begin{equation}
\label{eq:H0}
\tilde{H}_t^* = \frac{\mu}{\gamma\sigma^2}\ ;
\end{equation}
the amount $\tilde H_t = H_tS_t = \frac{\mu}{\gamma\sigma^2}$ is commonly referred to as \textit{the Merton line}, which gives the  value for c,
\begin{equation}
\label{eq:c0}
    c = \frac{2\sigma^2}{2\rho\sigma^2+\mu^2}\ .
\end{equation}
It is well-known and easy to verify that the solution to \eqref{eq:HJBmerton} is the equal to the optimal value function in \eqref{eq:valueFunction0}.

\subsection{Trading with Execution Costs (Case $\epsilon > 0$)}
\label{sec:tex}
Perturbation theory comprises mathematical methods for finding an approximate solution to a problem, by starting from the exact solution of a related, simpler problem. For the problem at hand, we would like to perturb the system from the simple case of no transaction costs to the case with proportional transaction costs. With the introduction of these costs to the value function, the optimal solution will tend to deviate from the value found in the case of a Merton problem as discussed in the previous section. This cost acts as a constraint against re-balancing, which means that, to maintain an optimal portfolio at each time instant, $t$ the investor needs to forgo an amount equal to the execution cost. 

The inclusion of transaction costs makes it important to keep track of the changes in the number of assets held in the portfolio because any change through buying/selling in the market incurs a cost to the investor. The additional cost of re-balancing here moves the optimal away from the ideal solution as obtained in the Black-Scholes world. Hence changes in the number of risky assets, now govern the solution at each time $t$, hence we define control variable to be $h_{t}$ rather than $H_t$. However, small-$\epsilon$ behavior allows for the optimal solution to stay close to the Merton line, or as it is referred to in the work of \cite{garleanuPedersen}, to maintain closeness to the ``aim portfolio''.

Following the dynamic programming approach as seen in the case for $\epsilon = 0 $, we write the HJB equation for the optimal control problem for the variable $h_{t}$ conditioned on values of the state variables at time $t$. Based on \eqref{eq:wpnl}, the value function in \eqref{eq:valueFunction} is the solution to the HJB equation,
\begin{align}
\nonumber
-\frac{1}{\gamma} e^{-\gamma w} - \rho V +\frac{\sigma^2s^2}{2}V_{ss}+\frac{\sigma^2\xi^2s^2}{2}V_{ww}+\mu sV_s+\mu\xi s V_w+\xi\sigma^2s^2V_{ws}&\\
\label{eq:HJBeps}
+\max_h\left(hV_\xi-\frac{\epsilon}{2}s h^2V_w\right)&=0\ ,
\end{align}
for $(w,s,\xi)\in\mathbb R\times\mathbb R^+\times\mathbb R^+$. We get \eqref{eq:h1} as the value by using first order conditions to maximize the control equation  $h$ 
\begin{equation}
\label{eq:h1}
    h^* = \frac{V_\xi}{\epsilon s V_w}\ .
\end{equation}
Assuming an exponential utility function for the investor, as given by $U(w) = -e^{-\gamma w}/\gamma$ we can employ the ansatz, 
\begin{equation}
\label{eq:ans}
V(w,s,\xi) = -\frac{1}{\gamma} e^{-\gamma w} G(s,\xi)\ ,
\end{equation}
we are able to write $h^{*}$ in the following manner,
\begin{equation}
\label{eq:h}
h^{*} = \frac{V_{\xi}}{\epsilon s V_{w}} = -\frac{G_{\xi}}{\epsilon \gamma s G}.
\end{equation}
Substituting \eqref{eq:ans} into \eqref{eq:HJBeps}, the HJB equation is transformed into a PDE in the variable $(s,\xi)$, which we proceed subsequently to solve. With these definitions and assumptions we get the following differential equation to solve,
\begin{equation}
\label{eq:nonlinearG}
1 +\mathcal L^\xi G- \rho G-\frac{(G_\xi)^2}{2\gamma\epsilon sG}=0\ ,
\end{equation}
for $(s,\xi)\in\mathbb R^+\times\mathbb R^+$, where we have defined the operator,
\begin{equation}
\label{eq:Lop}
    \mathcal L^\xi \triangleq \frac{\sigma^2s^2}{2}\frac{\partial^2}{\partial s^2}+\left(\mu s-\gamma\xi\sigma^2s^2\right)\frac{\partial}{\partial s}  +\frac{\gamma^2\sigma^2\xi^2s^2}{2} -\gamma\mu\xi s  \ .
\end{equation}

\subsection{Expanding in $\sqrt\epsilon$ Power Series}
It can be seen that \eqref{eq:nonlinearG} is singular due to the presence of $\epsilon$ in the denominator. Following the approach in Chapter 4 of \cite{fouque2011multiscale}, in order to tackle this singularity we perturb the system with a small quantity $\sqrt\epsilon$ and formulate the solution to the new perturbed system by a series given by \eqref{eq: PertG}.
We insert the expansion \eqref{eq: PertG} into \eqref{eq:nonlinearG} and collect the order-$1/\epsilon$ terms to find,
\begin{equation}
\label{eq:Gxi0}
G_\xi^{(0)} = 0 \ .
\end{equation}
Inserting the expansion in the formula for control $h^*$ we can approximate the solution employing Taylor expansion for $G$,
\begin{equation}
\label{eq:tradingh}
h^* = \frac{V_\xi}{\epsilon s V_w}=- \frac{G_\xi}{\gamma\epsilon s G} = - \frac{G_\xi^{(1)}}{\gamma\sqrt\epsilon s G^{(0)}}+\frac{G_\xi^{(1)}G^{(1)}}{\gamma s(G^{(0)})^2}-\frac{G_\xi^{(2)}}{\gamma sG^{(0)}}+O(\sqrt\epsilon)\ ,
\end{equation}
Our task now is to calculate each of the terms appearing in \eqref{eq:tradingh} in order to approximate the value of $h^*$ at each time $t$, with a high order of accuracy. We thus proceed by grouping together coefficients of equal powers of $\sqrt\epsilon$.

\label{subsec:sqeps}
As summarized at the end of the previous section, as a first step we look at the coefficients of order 1. We refer to them as first-order terms,
\begin{equation}
\label{eq:order1terms}
1 -\rho G^{(0)} +\mathcal L^\xi G^{(0)} -\frac{(G_\xi^{(1)} )^2}{2\gamma sG^{(0)} }=0\ .
\end{equation}
We base our expansions around the function $G^{(0)}$ that is the solution to \eqref{eq:HJBmerton} from the Merton problem (i.e., the case $\epsilon=0$). As $G^{(0)}$ is constant with respect to $s$ and $G_\xi^{(0)} = 0$ is as shown in \eqref{eq:Gxi0}, we complete the square in \eqref{eq:order1terms} to get,
\begin{align}
\label{eq:G1equation} 
\frac12\left(\gamma\sigma s\xi-\frac{\mu}{\sigma}\right)^2G^{(0)}-\frac{\mu^2}{2\sigma^2}G^{(0)}+1 -\rho G^{(0)} -\frac{(G_\xi^{(1)} )^2}{2\gamma sG^{(0)} }&=0\ .
\end{align}
Here we use the value for $G^{(0)}$ as calculated in \eqref{eq:c0}, 
\begin{equation}
\label{eq:G0}
G^{(0)} =  \frac{2\sigma^2}{2\rho\sigma^2+\mu^2} \ ,
\end{equation}
and substitute into \eqref{eq:G1equation} to solve for $G_\xi^{(1)}$,
\begin{equation}
\label{eq:gxi1sol}
    G_\xi^{(1)}(s,\xi)=\pm\sqrt{\gamma s\left(\frac{\mu}{\sigma}- \gamma \sigma s\xi \right)^2}G^{(0)}\ .
\end{equation}
The sensible root to take here is,
\begin{equation}
\label{eq:Gx1}
G_\xi^{(1)}(s,\xi)=-\sqrt{\gamma s}\left(\frac{\mu}{\sigma}-\gamma \sigma s\xi\right)G^{(0)}\ .
\end{equation}
By choosing the negative sign in \eqref{eq:gxi1sol} we have selected a specification under which the optimal amount of wealth in the risky asset will be mean-reverting towards the Merton portfolio given by \eqref{eq:H0}. A deviation from the Merton line is corrected by changing the number of assets in the portfolio, with the optimal change geing the amount $h^{*}$ given by equation $\eqref{eq:h}$. When $
\xi < \mu/(\gamma \sigma^2 s)$ then the portfolio is below the Merton line, and so the value of $G^{(1)}_{\xi}$ should be negative so that $h^{*}$ is a positive quantity, i.e., $h^*$ positive will increase the portfolio's holdings in the risky asset. For the opposite case when $
\xi > \mu/(\gamma \sigma^2 s)$, the portfolio is above the Merton line and the change $h^*$ will be a negative quantity so as to reduce the portfolio's holdings in the risk asset back down to the Merton line.

The function $G^{(1)}$ is found by integrating \eqref{eq:Gx1} with respect to $\xi$, after which we need to look for a function $C(s)$ that is constant in $\xi$ (i.e., a constant of integration), which leads to the following general solution for $G^{(1)}$ where $G^{(0)}$ is given in \eqref{eq:G0}, 
\begin{align}
\label{eq:G1_general}
G^{(1)}(s,\xi) &=C(s)+\frac{\sqrt{\gamma^3\sigma^2 s}G^{(0)}}{2s}\left(\frac{\mu}{\gamma\sigma^2}-\xi s\right)^2\ .
\end{align}

\subsection{Explicit Expression for $C(s)$}
\label{sec:epsterm}
Looking at the order $\sqrt\epsilon$ terms in the PDE we find,
\begin{align}
\label{eq:orderEpsTerms}
&\mathcal L^\xi G^{(1)}-\rho G^{(1)} -\frac{G_\xi^{(1)}}{2}\left(\frac{2G_\xi^{(2)}}{\gamma sG^{(0)} }-\frac{G_\xi^{(1)}G^{(1)}}{\gamma s(G^{(0)})^2}\right)=0\ .
\end{align}
Rearranging \eqref{eq:orderEpsTerms} gives us an explicit formula for $G_\xi^{(2)}$,
\begin{align}
\label{eq:G2}
G_\xi^{(2)}&= \gamma sG^{(0)}\frac{\mathcal L^\xi G^{(1)}-\rho G^{(1)}}{G_\xi^{(1)}}+\frac{G_\xi^{(1)}G^{(1)}}{2G^{(0)}}\ .
\end{align}
We shall use the formula for $G_\xi^{(2)}$ in \eqref{eq:G2} to infer properties of the solution for $G$, and also ensure that the solution obtained is well-behaved and stable. Ideally we do not expect significant deviations from the Merton solution due to the mean reversion towards the optimal.

It can be seen that formula \eqref{eq:G2} is singular along the aim portfolio's parameterized curve, 
\begin{equation}
\label{eq:mertxi}
    \xi(s) = \frac{\mu}{\gamma\sigma^2 s}\ ,
\end{equation}
for all $s>0$. We can see that $G_\xi^{(1)}(s,\xi(s))\equiv 0$ thus making it necessary for the numerator in the first term of \eqref{eq:G2} to approach zero for $\xi$ near to $\xi(s)$ in order to ensure that the equation is non-singular. 
Therefore, we look for $G^{(1)}$ along $\xi(s)$ such that the numerator in \eqref{eq:G2} is zero,
\begin{equation}
\label{eq:g1co}
\left(\mathcal L^\xi-\rho \right)G^{(1)}\Big|_{\xi=\xi(s)}=0\ .
\end{equation}
Recalling the definition of $\mathcal L^\xi$ as given in \eqref{eq:Lop}, we calculate the various partial differential terms involved in \eqref{eq:g1co}, 
\begin{align}
\label{eq:Ccomp}
\nonumber
G^{(1)}(s,\xi(s))&=C(s)\\
G_s^{(1)}(s,\xi)\Big|_{\xi=\xi(s)}&=C_s(s)\\
\nonumber
G_{ss}^{(1)}(s,\xi)\Big|_{\xi=\xi(s)}&=C_{ss}(s)+\frac{\mu^2}{\sqrt{\gamma s}\sigma^3 s^2} G^{(0)} .
\end{align}
Using these results and $G^{(0)}=  2\sigma^2/ (2\rho\sigma^2+\mu^2) $, we then look for a solution of $C(s)$ that makes the numerator in the singular term in \eqref{eq:G2} go to zero along $\xi(s)$. Such a $C(s)$ is the solution to the following ODE derived from \eqref{eq:g1co} using the derivatives from \eqref{eq:Ccomp}, 
\begin{align}
\label{eq:Cpde}
\frac{\sigma^2s^2}{2}C_{ss}-\left(\rho+\frac{\mu^2}{2\sigma^2}\right)C+\frac{\mu^2}{2\sqrt{\gamma s}\sigma } G^{(0)} =\left(\mathcal L^\xi-\rho\right) G^{(1)}\Big|_{\xi=\xi(s)}&=0\ .
\end{align}
The solution to \eqref{eq:Cpde} is found by applying the change of variables $x=\log(s)$, upon which an equation for $\widetilde C(x) = C(e^x)$ can be rewritten as a 2nd-order linear non-homogeneous equation with constant coefficients,
\begin{equation}
\label{eq:Cpdex}
    \frac{\sigma^2}{2}\widetilde C_{xx}-\frac{\sigma^2}{2}\widetilde C_x-\left(\rho+\frac{\mu^2}{2\sigma^2}\right)\widetilde C+\frac{\mu^2}{2\sqrt{\gamma }\sigma } G^{(0)}e^{-\frac{1}{2}x}=0\ ,
\end{equation}
which via the superposition principle has general solution,
\begin{equation}
\label{eq:Csup}
    \widetilde C(x) = \widetilde C_h(x)+\widetilde C_p(x)\ ,
\end{equation}
where $\widetilde C(x)$ is a solution to the homogeneous part of \eqref{eq:Cpdex},
\begin{equation}
\label{eq:Ch}
    \widetilde C_h(x)=A_1e^{\lambda_+x}+A_2e^{\lambda_-x}\ ,
\end{equation}
where $\lambda_\pm = 1/2\pm\sqrt{1/4+(2\sigma^2\rho+\mu^2)/\sigma^4}$, and where $\widetilde C_p(x)$ is a particular solution to the inhomogeneous part of \eqref{eq:Cpdex},
\begin{equation}
\label{eq:Cpt}
\widetilde C_p(x)=
\begin{cases}
A_3e^{-\frac12x}\qquad\hbox{if $\lambda_-\neq -\frac12$} \\
A_3xe^{-\frac12x}\qquad\hbox{otherwise}\ .
\end{cases}
\end{equation}
For both cases the constant $A_3$ in \eqref{eq:Cpt} can be solved for,
\begin{equation}
\label{eq:A3}
A_3 = \frac{\mu^2}{2\sqrt{\gamma }\sigma } G^{(0)}
   \times \begin{cases}
1/(
 \rho+\mu ^2 / 2 \sigma ^2 -3
   \sigma ^2/8)\qquad\hbox{if $\lambda_-\neq -\frac12$}\\
1/\sigma^2\qquad\qquad\qquad\hbox{otherwise}\ .
\end{cases}
\end{equation}
It remains to find the constants $A_1$ and $A_2$ in \eqref{eq:Ch}. If we assume the homogeneous solution is independent of the risk aversion rate, then $C_h(x)$ is the solution to \eqref{eq:Cpdex} as $\gamma\rightarrow \infty$, which is the case of infinite risk aversion wherein the investor has zero allocation in the risky asset, and hence $\lim_{\gamma\rightarrow \infty}G^{(1)}(e^x,\xi(e^x))=C_h(x)=0$, or $A_1=A_2 = 0$. Under this assumption, the particular solution as given in \eqref{eq:A3} is all that remains, and we are able to write \eqref{eq:Csup} in terms of the original variable,
\begin{equation}
\label{eq:Css}
    C(s) =
 \frac{\mu^2}{2\sqrt{\gamma s }\sigma } G^{(0)}\times \begin{cases}
1/(
 \rho+ \mu ^2 / 2 \sigma ^2 -3
   \sigma ^2 /8)\qquad\hbox{if $\lambda_-\neq -\frac12$}\\[10pt]
\log(s)/\sigma^2 \qquad\qquad\qquad\hbox{otherwise}\ .
\end{cases}
\end{equation}
Inserting the expression of \eqref{eq:Css} into \eqref{eq:G1_general}, and noticing that $\lambda_- = -\frac12$ if and only if $4\mu^2=3\sigma^4-8\sigma^2\rho$, we have $G^{(1)}$ given explicitly as,
\begin{align}
\nonumber
G^{(1)}(s,\xi)& = \frac{\sqrt{\gamma^3\sigma^2 }G^{(0)}}{2\sqrt s}\left(\frac{\mu}{\gamma\sigma^2}-\xi s\right)^2\\
\label{eq:G1}
&~+ \frac{\mu^2}{2\sqrt{\gamma  s}\sigma } G^{(0)}\times \begin{cases}
1/(
 \rho+ \mu ^2/2 \sigma ^2 - 3
   \sigma ^2/8)\qquad\hbox{if $|\mu|\neq\frac12\sqrt{3\sigma^4-8\sigma^2\rho}$}\\[10pt]
\log(s)/\sigma^2\qquad\qquad\qquad\hbox{otherwise}\ .
\end{cases}
\end{align}

\subsection{Higher-Order Analysis}
\label{sec:explG}
We shall now take up the analysis of $G^{(2)}_{\xi}$ and work out an explicit expression for the same. We begin with \eqref{eq:G2}. A preliminary glance at the expression shows us that there is a need for close examination of the expansion near the Merton line as $\xi(s) \rightarrow \mu/\gamma \sigma^2 s$. This is due to the fact that there is $G_\xi^{(1)}$ in the denominator of the first term of \eqref{eq:G2}. We can rearrange \eqref{eq:G2} so that the potentially singular terms are grouped together,
\begin{equation}
\label{eq:criticalQuantity}
G_\xi^{(2)} - \frac{G_\xi^{(1)}G^{(1)}}{2G^{(0)}}=\gamma sG^{(0)}\frac{\mathcal L^\xi G^{(1)}-\rho G^{(1)}}{G_\xi^{(1)}}, 
\end{equation}
and from here we analyze whether or not the right-hand side will destabilize the expansion at the Merton line. We proceed by factoring the right-hand side of \eqref{eq:criticalQuantity} into multiples of $(\xi s - \mu/\gamma \sigma^2)$. Using the expression for $G^{(1)}$ from \eqref{eq:G1}, for the non-critical case $|\mu|\neq \frac12\sqrt{3\sigma^4-8\rho\sigma^2}$ we have the following expression for $G_\xi^{(2)}$,
\begin{equation}
\label{eq:Gxi2explicit}
G_\xi^{(2)}(s,\xi) = \left(\xi s - \frac{\mu}{\gamma \sigma^2}\right) \Big(D_1(s,\xi) +  D_2\Big) + \frac{G_\xi^{(1)}(s,\xi)G^{(1)}(s,\xi)}{2G^{(0)}}\ ,
\end{equation}
where the coefficients in the formula of \eqref{eq:Gxi2explicit} are
\begin{align}
\nonumber
D_1(s,\xi) &=\frac{\sigma ^2 \left(\gamma ^2 \left(4 \mu +8 \rho
   -3 \sigma ^2\right)-4 \gamma ^4 \xi ^2 s^2
   \sigma ^2\right)}{8 \gamma  \left(\mu ^2+2 \rho 
   \sigma ^2\right)}\\
   \nonumber
   &\hspace{2cm}-\frac{\left(32 \mu ^2 \gamma^2\sigma^2-4
   \gamma ^3 \xi  s \left(2 \mu +3 \sigma
   ^2\right)\right)}{8 \gamma  \left(\mu ^2+2 \rho 
   \sigma ^2\right)\left(4
   \mu ^2+8 \rho  \sigma ^2-3 \sigma ^4\right)}\\
  \label{eq:explicitD1andD2}
D_2 &= -\frac{\mu  \sigma ^4 \left(\gamma ^2\left(4 \mu ^2+8 \rho  \sigma ^2-3
   \sigma ^4\right)+4 \mu 
   \gamma^2\sigma^2\right)}{\gamma  \left(\mu ^2+2
   \rho  \sigma ^2\right)\left(4 \mu ^2+8 \rho  \sigma ^2-3
   \sigma ^4\right)}\ .
\end{align}
The explicit expression for $G_\xi^{(2)}$ given by equations \eqref{eq:Gxi2explicit} and \eqref{eq:explicitD1andD2} is a somewhat pleasant surprise because higher-order terms are often hard to find in exact form. For the critical case of $|\mu|=\frac12\sqrt{3\sigma^4-8\rho\sigma^2}$, the expression for $G_\xi^{(2)}$ does not have the quadratic form as in the non-critical case, but factoring into multiples of $\xi s - \mu/\gamma \sigma^2 $ yields a stable expression for $G_\xi^{(2)}$ of the form $G_\xi^{(2)} = D(s,\xi) + G_\xi^{(1)}G^{(1)}/{2G^{(0)}}$, where $D(s,\xi)$ is non-singular for for $\xi = \xi(s)$. This completes the derivation for the expansion for $G_\xi^{(2)}$.\\

To summarize, we have derived the first two terms in the expansion and obtained an approximation,
\begin{equation}
\label{eq:mainExpansion}
V(w,s,\xi)\approx U(w)\left(G^{(0)}+\sqrt\epsilon G^{(1)}(s,\xi)\right)\ ,
\end{equation}
where $G^{(0)}$ and $G^{(1)}$ are given by equations \eqref{eq:G0} and \eqref{eq:G1}. We have also calculated the next term $G_\xi^{(2)}$ to ensure stability at order $\epsilon$ in the expansion, and also we have a proposed approximate control if we insert these terms in \eqref{eq:tradingh}.

\section{Verification Lemmas and Accuracy of Expansion}
\label{sec:verificationAndExistence}
This section proves the verification lemma for the solution of HJB equation \eqref{eq:HJBeps}, and then shows accuracy of the expansion. We first do the proof for the case when a classical solution exists, and then second for the case where there is only known to be a non-smooth viscosity solution. Using the separation of variables of $V(w,s,\xi) =U(w)G(s,\xi)$, we are able to show the existence of a viscosity solution to \eqref{eq:nonlinearG} for the value function $G$.

\subsection{Viscosity Approach to Verification}
\label{sec:verificationLemma}

Equation \eqref{eq:nonlinearG} is the HJB equation for the following optimization problem,
\begin{align}
\label{eq:dualProblem}
G(s,\xi)&=\inf_{h\in\mathcal A}\mathbb E\int_0^{\infty} e^{- \rho t}e^{- \gamma \int_0^t\left(H_udS_u-\frac{\epsilon}{2}S_uh_u^2du\right) }dt\ ,
\end{align}
where $\mathcal A$ is the same admissible set as defined in \eqref{eq:A}. By examination of the objective in \eqref{eq:dualProblem}, one can see that a finite value is obtained with a strategy that abstains entirely from trading the risky position initially and then sets $H_t\equiv H_0$ and $h_t\equiv0$ on $[0,\infty)$. Therefore $G(s,\xi)\leq \frac{1}{\rho}e^{\gamma s\xi}$ for all $(s,\xi)$, and there is the following general bound on the optimal $G$,
\begin{equation}
\label{eq:Gbound}
\frac{1}{\rho+\mu^2/2\sigma^2}\leq G(s,\xi)\leq  \frac{1}{\rho}e^{\gamma s\xi}<\infty\ ,
\end{equation}
for any point $(s,\xi)$. 

Let $G$ be a classical solution to \eqref{eq:nonlinearG}, and for any admissible $h$ let $(\tau_k)_{k=1,2,3,\dots}$ be a family of increasing stopping times with $\tau_k\rightarrow \infty $ almost surely. From It\^o's lemma there is the following bound,
\begin{align}
\label{eq:GStopBound}
G(s,\xi)&\leq\mathbb E\int_0^\infty e^{- \rho t}e^{- \gamma \int_0^t\left(H_udS_u-\frac{\epsilon}{2}S_uh_u^2du\right) }dt+\overline{\lim}_k\mathbb E\left[e^{-\tau_k\rho }G(S_{\tau_k},H_{\tau_k})\right]\ ,
\end{align}
which proves the verification if $\overline{\lim}_k\mathbb E\left[e^{-\tau_k\rho }G(S_{\tau_k},H_{\tau_k})\right] = 0$ \citep[see][page 139, Chapter 3]{flemingSoner2006}. However, we don't know that there exists a classical solution and we don't know that the limit supremum equals to zero. Nonetheless, we can still prove a verification for viscosity solutions, which is useful because it shows that our $\sqrt\epsilon$ expansion is an approximation of an optimal viscosity solution.\footnote{A viscosity super solution $G(s,\xi)$ of \eqref{eq:nonlinearG} is one where for any smooth test function $\phi(s,\xi)$ and any point $(s_0,\xi_0)$ such that $G(s_0,\xi_0) = \phi(s_0,\xi_0)$ and $G(s,\xi) \geq \phi(s,\xi) $ for all other $(s,\xi)$, we have $
    -\left(1-\rho \phi +\mathcal L^\xi \phi-\frac{\phi_\xi^2}{2\epsilon\gamma s\phi}\right)\geq 0$. A viscosity sub solution is one where the inequalities are reversed. Simply stating that $G$ is a \textit{viscosity solution} means that it is both a viscosity sub and super solution \citep[see][]{crandall1987,flemingSoner2006}.} The PDE operator $-(1-\rho \phi +\mathcal L^\xi \phi-\phi_\xi^2/2\epsilon\gamma s\phi)$ is degenerate elliptic and so a classical solution of \eqref{eq:nonlinearG} is also a viscosity solution, see \cite{crandall1987,flemingSoner2006}, and so for verification it is sufficient to show the inequality of \eqref{eq:GStopBound} for viscosity solutions.

\subsection{The Expansion's Accuracy}
 We now present an argument to show that this unique viscosity solution is approximated by the $\sqrt\epsilon$ expansion of $G=  G^{(0)}+\sqrt\epsilon G^{(1)}+\mathcal O(\epsilon)$. The expansion indicates that the optimal $H_t$ stays close to the Merton line $\xi(s) = \mu/(\gamma\sigma^2s)$, and from the optimal strategy approximation of \eqref{eq:tradingh} there is the following admissible but sub-optimal strategy to consider,
\begin{equation}
\label{eq:hForTrading}
    \hat h =-\frac{G_\xi^{(1)}}{\gamma s\sqrt\epsilon G^{(0)}}= \frac{1}{\sqrt{\epsilon\gamma s}}\left(\frac{\mu}{\sigma}-\gamma \sigma s\xi\right)\ ,
\end{equation}
which is mean reverting with a $1/\sqrt\epsilon$ rate similar to the strategies shown in \cite{voss2016} and \cite{rogers2010}. Following the same method of proof as done in \cite{rogers2010}, we use strategy $\hat h$ given by \eqref{eq:hForTrading} to derive linear equations, for which the linear theory for singularly-perturbed PDEs applies, and hence there will be an order-$\sqrt\epsilon$ bound on the error between $G^{(0)}$ and the optimal value function.

We add an exit criterion to the strategy where for an arbitrarily large (but finite) parameter $M$ we cease trading if $S_t\notin (1/M, M)$ or if $|\mu/(\gamma\sigma^2)-H_tS_t|\geq M$. Defining the open set $\mathcal D =\left\{(s,\xi)\in\mathbb R^+\times\mathbb R^+:1/M<s<M~~\hbox{and}~~\left|\mu/(\gamma\sigma^2)-s\xi\right|<M\right\}$, and defining the stopping time $\tau = \inf\{t>0: (S_t,H_t)\notin \mathcal D\}$, we can define another value function that is equal to the expected utility of the exiting strategy,
\begin{equation}
\label{eq:gSubOptimal}
    g(s,\xi) = \mathbb E\left[\int_0^\tau  e^{-\rho t}e^{-\gamma \int_0^t\left(H_udS_u-\frac{\epsilon}{2}S_uh_u^2du\right)}dt+e^{-\rho\tau}R(S_\tau,H_\tau)\right]\Bigg|_{h=\hat h}\ ,
\end{equation}
where $R(s,\xi) = e^{\gamma s\xi}\mathbb E\int_0^\infty e^{-\rho t-\gamma \xi S_t}dt$ is the value function evaluated for the admissible control of $h_t\equiv 0$. The value function $g$ defined in \eqref{eq:gSubOptimal} is a bound for the optimal,
\begin{equation}
\label{eq:subOptimalBound}
\frac{2\sigma^2}{2\sigma^2\rho+\mu^2}\leq G(s,\xi)\leq g(s,\xi)\ ,
\end{equation}
and is the Feynman-Kac representation of the solution to a linear PDE,
\begin{align}
\nonumber
1-\left(\rho+\frac{\mu^2}{2\sigma^2}\right) g + \frac{\sigma^2s^2}{2}g_{ss}+s\gamma\sigma^2\left(\frac{\mu}{\gamma\sigma^2} - \xi s\right)g_s  &\\
\nonumber
+ \gamma^2\sigma^2\left(\frac{\mu}{\gamma\sigma^2}-\xi s\right)^2 g+\sqrt{\frac{\gamma \sigma^2 }{\epsilon s}}\left(\frac{\mu}{\gamma\sigma^2}-\xi s\right)g_\xi&= 0\qquad\qquad(s,\xi)\in\mathcal D\\
\label{eq:linear_g}
g(s,\xi)&=R(s,\xi)\qquad(s,\xi)\in\partial\mathcal D\ .
\end{align}
Now because the coefficients in \eqref{eq:linear_g} are Lipschitz continuous and bounded on $\mathcal D$, it follows that the linear theory of \cite{fouque2011multiscale} applies, so we write the formal expansion $g(s,\xi) = g^{(0)}(s,\xi)+\sqrt \epsilon g^{(1)}(s,\xi)+\mathcal O(\epsilon )$. This expansion is valid for points $(s,\xi)$ away from $\partial\mathcal D$ so that there is a boundary layer to ensure accuracy as done in Chapter 5 of \cite{howison2005practical}, or how it is done for the free boundary in \cite{Fouque2001}. Carrying through the analysis we obtain the same approximation as that obtained in \cite{rogers2010},
\begin{equation}
\label{eq:gMbound}
g(s,\xi) \leq  g^{(0)}(s,\xi)+\sqrt\epsilon K^{M,\epsilon}(s,\xi)\ ,
\end{equation}
where $ K^{m,\epsilon}\geq 0$ is finite. Examination of the singular term in \eqref{eq:linear_g} leads to $g_\xi^{(0)} \equiv 0$, and so there is a constant zero-order term of $g^{(0)} \equiv 2\sigma^2/(2\sigma^2\rho+\mu^2) = G^{(0)}$. Moreover, examination of the order-1 terms leads to $g_\xi^{(1)}(s,\xi) = -\sigma\sqrt{\gamma^3s}\left(\frac{\mu}{\gamma\sigma^2}-\xi s\right)g^{(0)}$, which is solved to obtain $g^{(1)}(s,\xi) = c(s)+ \frac{\sigma}{2}\sqrt{\frac{\gamma^3}{s}}\left(\frac{\mu^2}{\gamma\sigma^2}-\xi s\right)^2 g^{(0)}$ with $c(s)$ solving the equation $\frac{s^2}{2}c''(s) - c(s)(2\sigma^2\rho+\mu^2)/2\sigma^4=0$; the general solution is $c(s) = c_1s^{\lambda_+}+c_2s^{-\lambda_-}$ where $\lambda_\pm =\frac{1}{2}\left(1 \pm \sqrt{1+4(2\sigma^2\rho+\mu^2)/\sigma^4}\right)$. Hence, using the bounds of equations \eqref{eq:subOptimalBound} and \eqref{eq:gMbound} we find closeness of the optimal value function $G(S,\xi)$ to the zero-order term in the expansion in the sub-optimal value function in \eqref{eq:gSubOptimal},
\begin{equation}
\label{eq:errorBound}
\frac{2\sigma^2}{2\sigma^2\rho+\mu^2}\leq G(s,\xi)\leq g(s,\xi)\leq  \frac{2\sigma^2}{2\sigma^2\rho+\mu^2}+\sqrt\epsilon  K^{M,\epsilon}(s,\xi)\ .
\end{equation}
for $(s,\xi)$ away from $\partial \mathcal D$.

A higher-order error bound beyond that given in \eqref{eq:errorBound} can be computed; it will have a term of $\mathcal O(\epsilon)$ plus an error term that will depend on $M$. The calculations to obtain the higher-order error are significantly more involved and so we do not show them here, but comments are in order to provide insight to the order-$\sqrt\epsilon$ expansion's accuracy. The expansion does not approximate well for $s$ near zero because $G^{(1)}$ dominates with $1/\sqrt s$ behavior. For points $(s,\xi)$ with $s\gg 1/M$ the expansion is a good approximation because for $\mu> 0$ the process $S_t$ stays away from $1/M$ with high probability. Another limitation is that the expansion loses accuracy for points $(s,\xi)$ far away from the Merton line given by \eqref{eq:mertxi}, which can be seen when consideration is given to $G^{(1)}$'s quadratic behavior in $\xi$, which will dominate the error terms when $\xi$ strays far from $\xi(s)$. However, the strategy $\hat h$ has an $H_t$ process that stays close to $\xi(S_t)$ with high probability when $\epsilon$ is small. The simulations of Section \ref{sec:monteCarlo} will demonstrate the points of this paragraph.

\section{Monte Carlo Simulations}
\label{sec:monteCarlo}
We test the performance of our results by analyzing the evolution of the wealth process, keeping the Merton problem as a benchmark.
For our simulation we begin the analysis by choosing the following values for the parameters, $S_0 = 100$, $W_0 = 100$, $\gamma  = 0.01$, $\sigma  = 0.15$, $\mu  = 0.05$, $T = 10$, $\epsilon  = 0.01$, and $\Delta t  = 0.004$ being the discrete time step used in a forward Euler scheme for simulating the SDEs. The choice of the time step will need to be smaller as $\epsilon$ gets small, so we make sure that $\Delta t$ is small enough to ensure reasonable order of variance in the transitions of the processes' discrete paths.

\begin{figure}[h]
\centerline{\includegraphics[width=80mm]{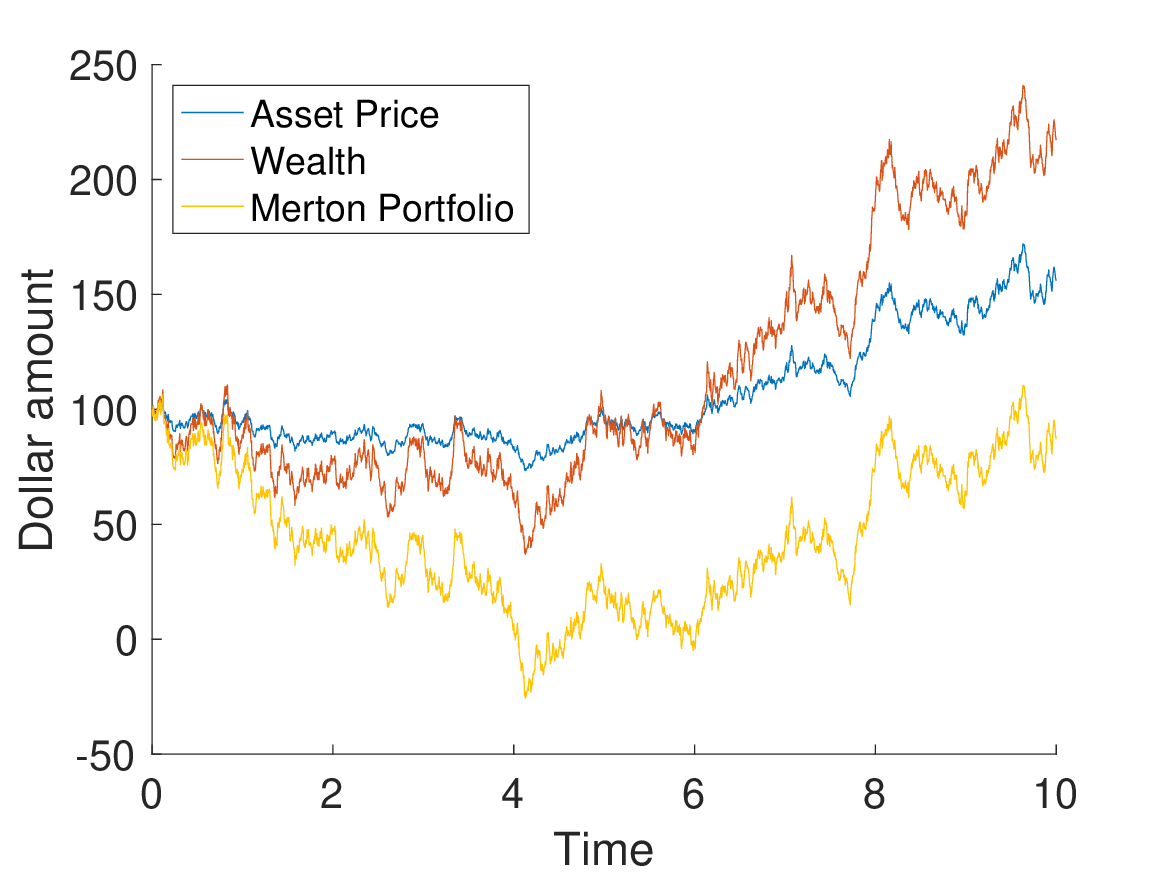}}
\vspace*{8pt}
\caption{The wealth process of the aim portfolio, and for comparison, shown alongside the asset price and the portfolio using the Merton strategy, with the following parameters, $\sigma  = 0.15$, $\mu  = 0.05$, $\epsilon  = 0.01$, and $\gamma  = 0.01$. The portfolio using Merton strategy has transaction costs removed from returns, hence it does not perform as well as the aim portfolio.}
\label{fig:wamer}
\end{figure}

We start by simulating the asset price $S_t$ using a standard normal random variable.  Equation \eqref{eq:wpnl} gives us the change in the wealth of the investor. In Figure \ref{fig:wamer} we first look at the variations of wealth process with the evolution of asset price and Merton portfolio as given by \eqref{eq:Wealth0} with transaction costs accounted for.

\begin{figure}[h]
\centerline{\includegraphics[width=80mm]{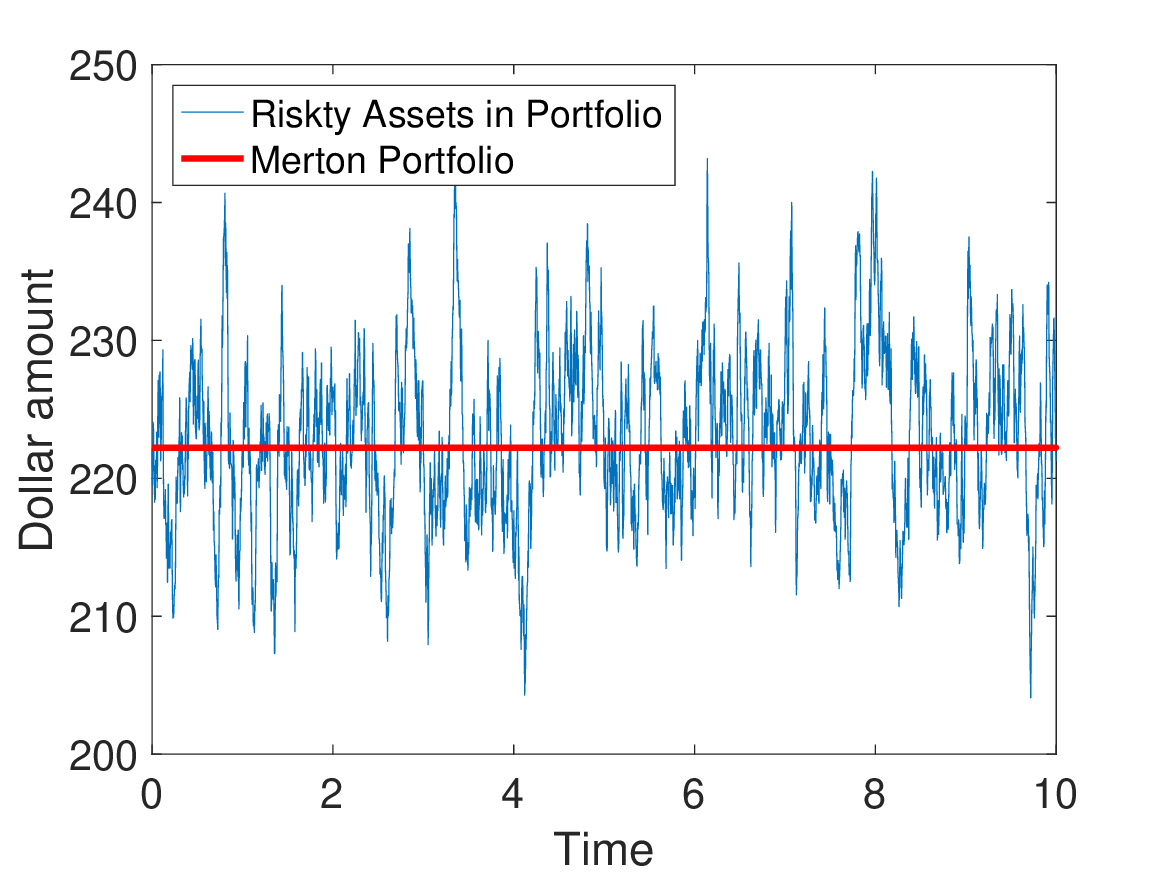}}
\vspace*{8pt}
\caption{Dollar amount of risky assets in the portfolio, $\tilde{H}$ shows mean reversion around the Merton portfolio. For small $\epsilon$ there is fast mean reversion to the Merton line. The plot shown here is for the following parameters, $\sigma  = 0.15$, $\mu  = 0.05$, $\epsilon  = 0.0001$, and $\gamma  = 0.01$.}
\label{fig:delaset}
\end{figure}

The $h$ thus determines the amount by which the number of assets held in the portfolio changes (discrete difference) from time $t$ to $t+\Delta t$ according to \eqref{eq:tradingh}. We plot \eqref{eq:H0} and $\tilde{H}$ to observe the mean reversion around the Merton portfolio, as shown in Figure \ref{fig:delaset}

\begin{figure}[h]
\centering
\subfloat[$\gamma$ = 0.01]{\includegraphics[width=50mm]{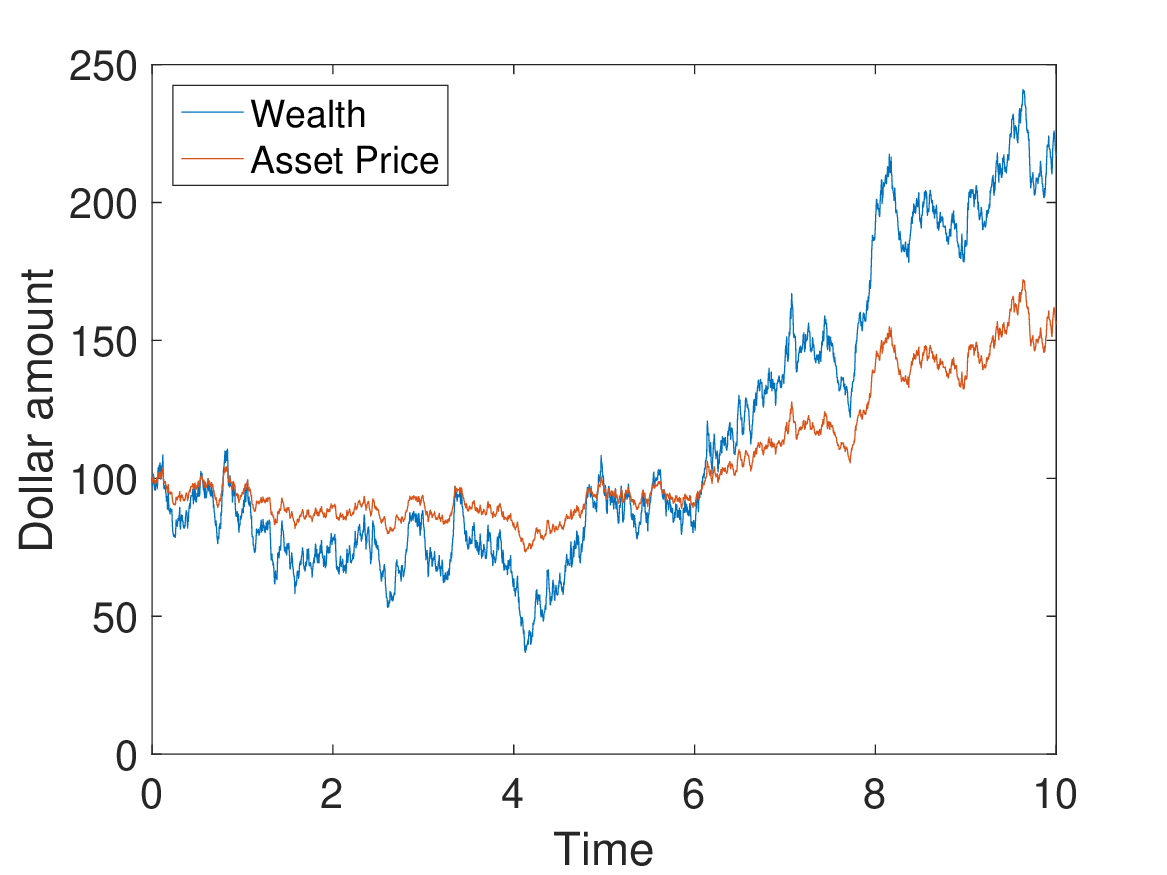}}
\subfloat[$\gamma$ = 0.02]{\includegraphics[width=50mm]{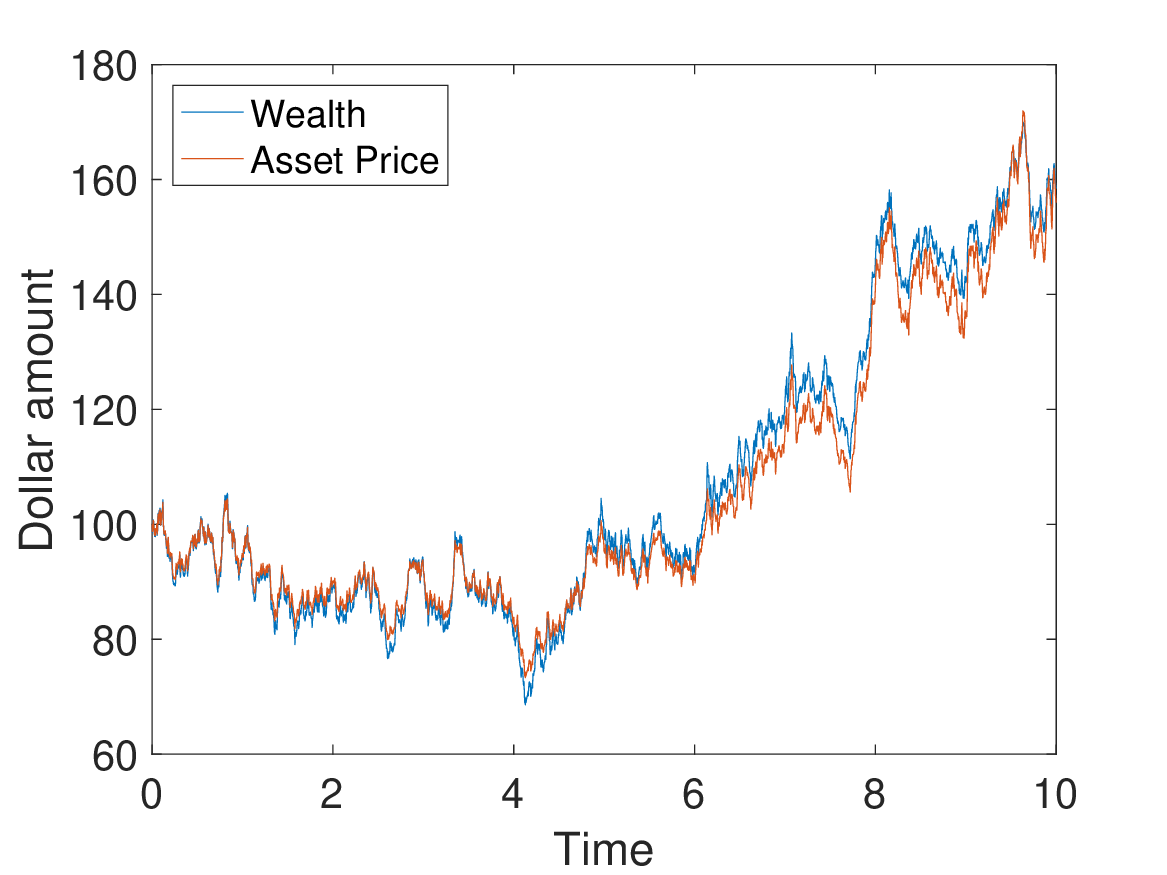}}
\hspace{0mm}
\subfloat[$\gamma$= 0.05]{\includegraphics[width=50mm]{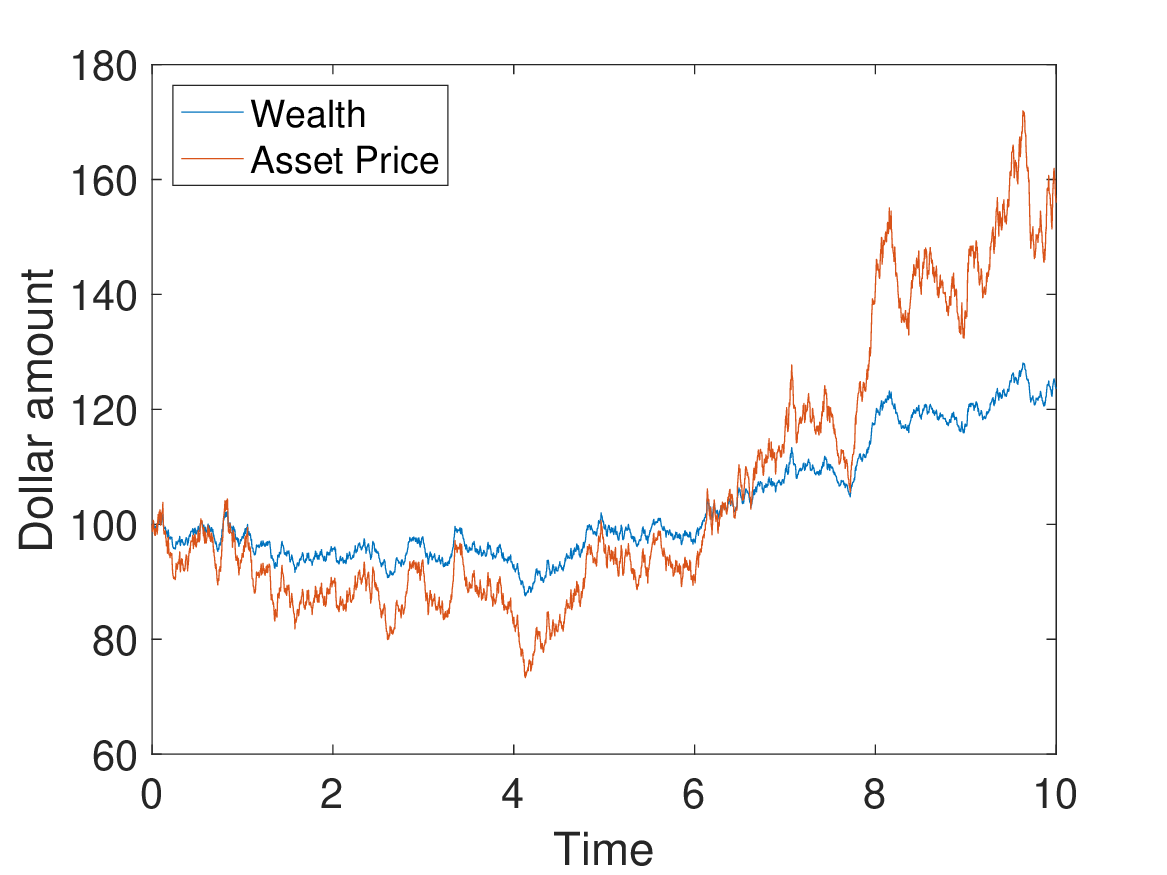}}
\subfloat[$\gamma$ = 0.1]{\includegraphics[width=50mm]{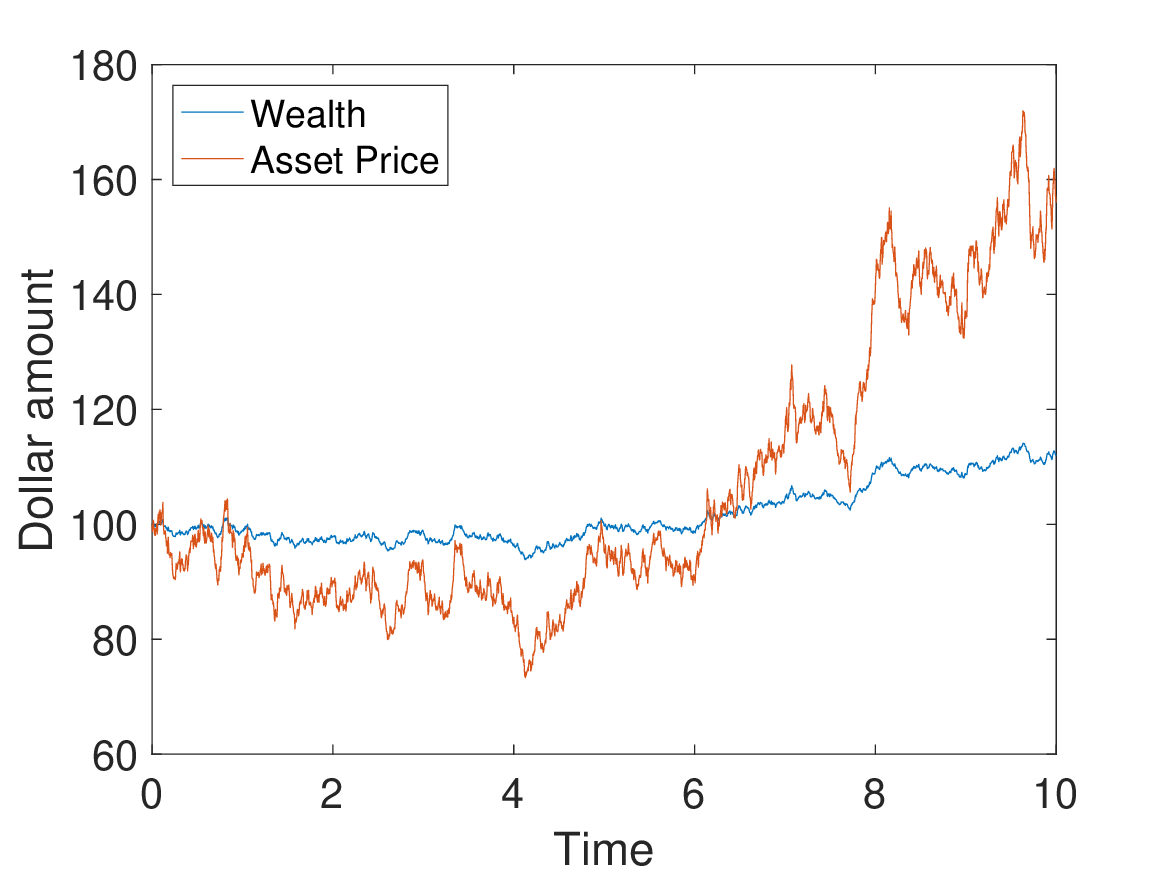}}
\caption{Evolution of the wealth process of an investor with $S_0 = 100$, $W_0 = 100$, $\sigma  = 0.15$, $\mu  = 0.05$, $\Delta t  = 0.004$, $T = 10$  and $\epsilon  = 0.01$. The subplots show the variation of wealth with respect to the stock price for different cases of risk aversion of investor. We see that for low risk aversion $\gamma = 0.01$, the wealth process has volatility that is in sync with the movements of the stock price. Higher values of $\gamma$ show lesser volatility in the wealth process compared to the asset price.}
\label{fig:fgamma}
\end{figure}

\begin{figure}[h]
\centering
\subfloat[$\epsilon$ = 0.0001]{\includegraphics[width=50mm]{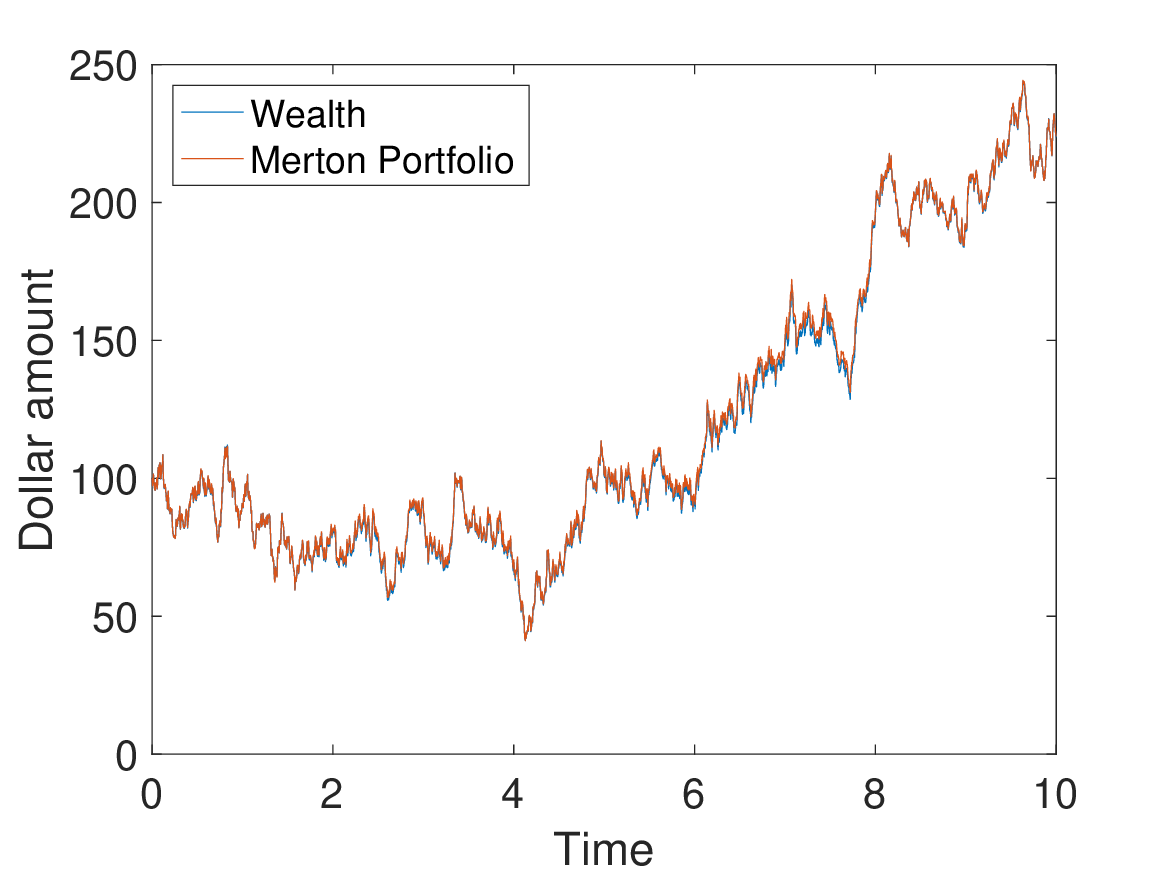}}
\subfloat[$\epsilon$ = 0.001]{\includegraphics[width=50mm]{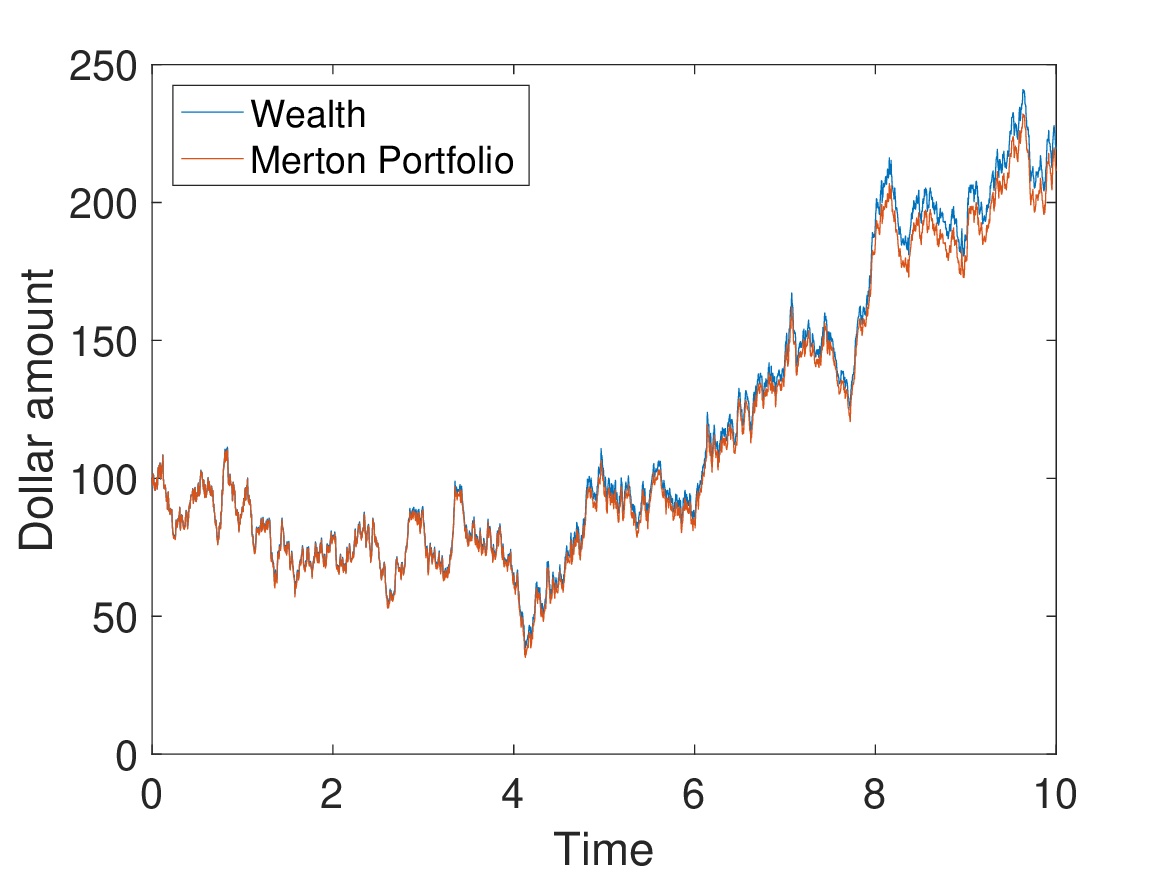}}
\hspace{0mm}
\subfloat[$\epsilon$= 0.005]{\includegraphics[width=50mm]{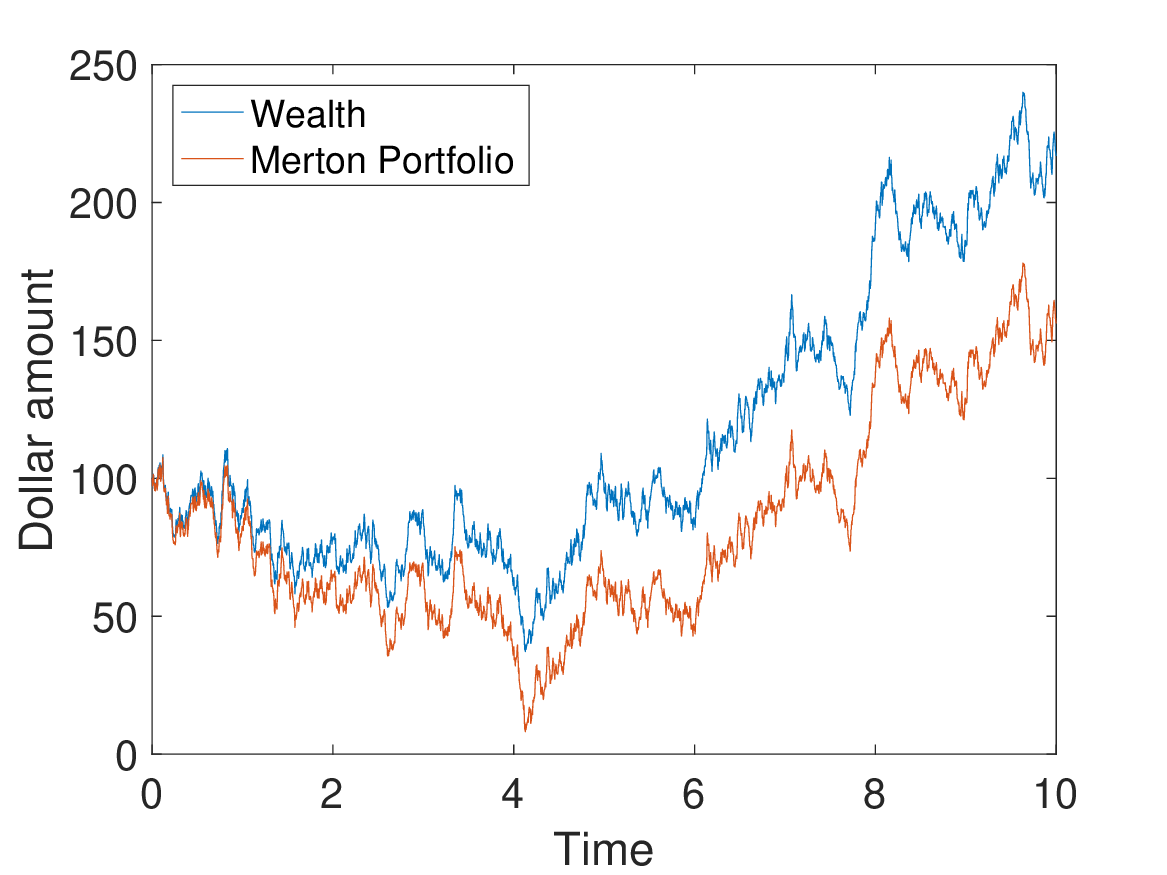}}
\subfloat[$\epsilon$ = 0.01]{\includegraphics[width=50mm]{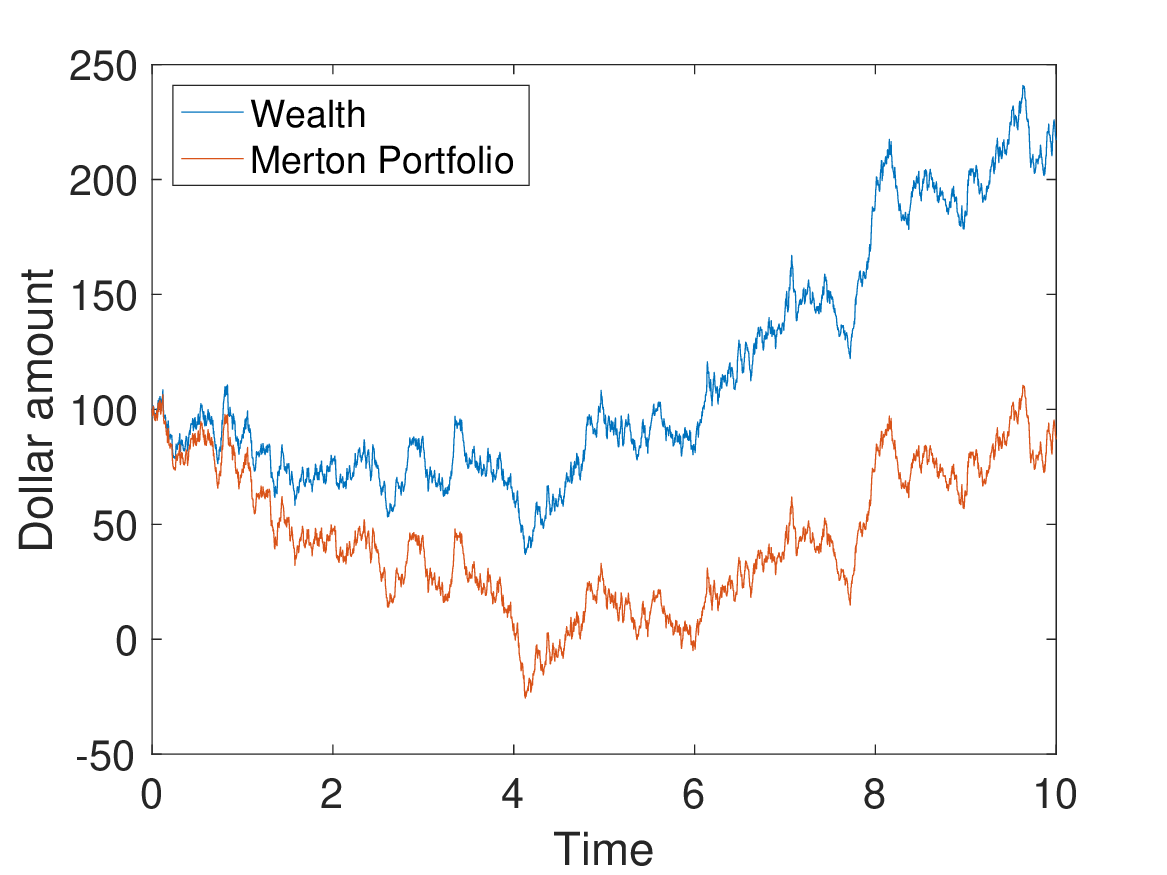}}
\caption{Evolution of the wealth process of an investor with $S_0 = 100$, $W_0 = 100$, $\sigma  = 0.15$, $\mu  = 0.05$, $\Delta t  = 0.004$, $T = 10$ and $\gamma  = 0.01$. The subplots show the variation of wealth with respect to the evolution of Merton portfolio for different $\epsilon$ values. We see that for low values of $\epsilon$, the wealth process moves closely with the Merton portfolio. Higher values of $\epsilon$ show higher deviation in the value of wealth over time with respect to the wealth of the Merton portfolio.}
\label{fig:fepsilon}
\end{figure}

We would also like to illustrate the variations of the wealth process with changing values of parameters $\epsilon$, which determines the magnitude of the trading penalty, and also with changing $\gamma$ which describes the risk aversion of the investor. Figure \ref{fig:fgamma} depicts the effect of $\gamma$ on the wealth. We can see that with lower risk aversion the investor's wealth fluctuates with the changing stock price and matches the volatility of prices. However with higher risk aversion the investor's wealth has less volatility, and this volatility decreases noticeably as we increase the value of $\gamma$. Similarly, we can show that the variation in $\epsilon$ helps achieve different behavior in the wealth process. As seen in Figure \ref{fig:fepsilon}, the wealth process tracks the Merton portfolio for smaller values of $\epsilon$, and has noticeable tracking error for higher values of $\epsilon$. 

\begin{figure}[h]
\centerline{\includegraphics[width=80mm]{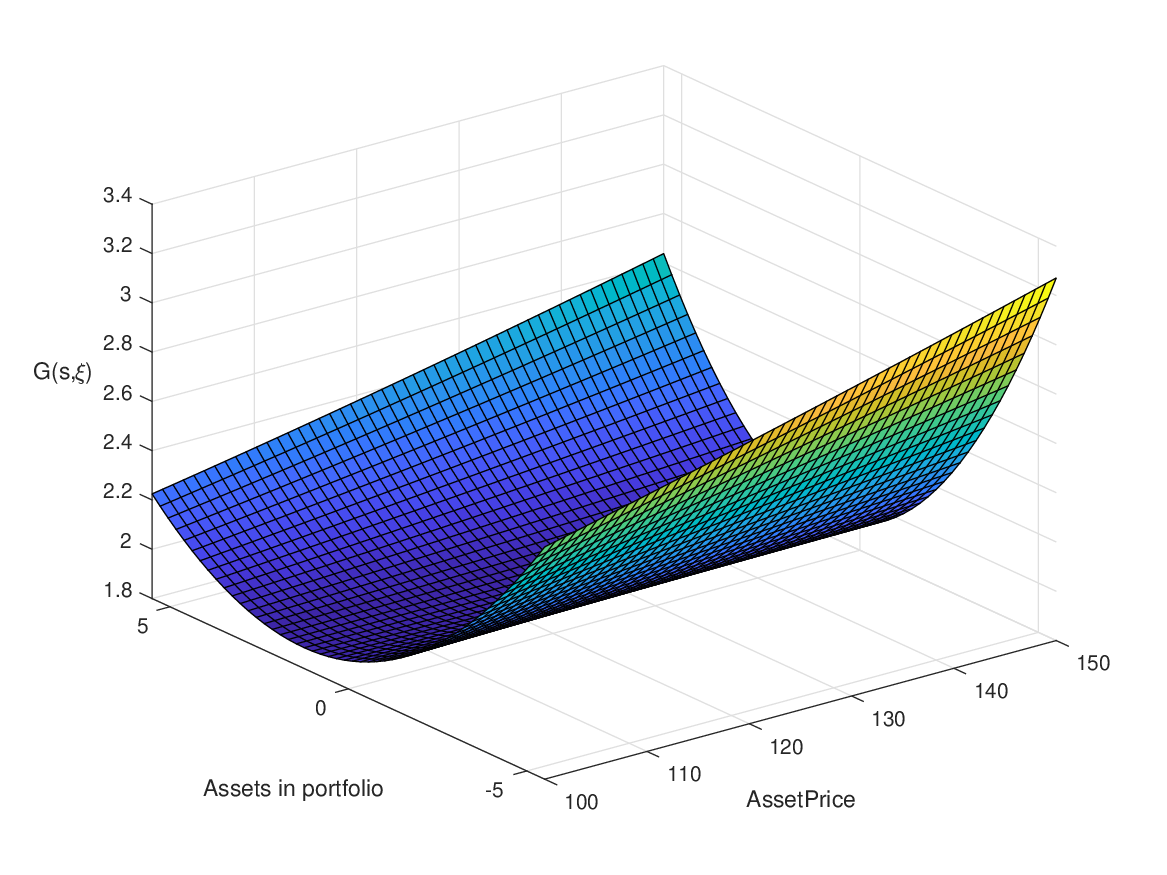}}
\vspace*{8pt}
\caption{Surface plot of the value function for variations in number of assets in the portfolio, $\xi$ and the price of the risky asset, $s$ for the following parameters, $\sigma  = 0.2$, $\mu  = 0.05$, $\epsilon  = 0.01$ and $\gamma  = 0.01$.}
\label{fig:invvalue}
\end{figure}

Finally, we also look at the effects of variation in asset price, $S$ and holdings, $\xi$ on the expansion $G$ at time t=0.004. The surface plot in Figure \ref{fig:invvalue} shows the approximated function $G = G^{(0)} + \sqrt\epsilon G^{(1)} +\mathcal O(\epsilon)$ which is as given in \eqref{eq:mainExpansion} without the effects of the exponential utility function, with $S$ and $\xi$ as the input variables. In particular, Figure \ref{fig:invvalue} shows how the value function decreases quadratically as $\xi$ deviates from the Merton line.

\begin{thebibliography}{}

\bibitem[Almgren \& Chriss, 2001]{almgren2001optimal}
Almgren, R. \& Chriss, N. (2001).
\newblock Optimal execution of portfolio transactions.
\newblock {\em Journal of Risk}, 3:5--40.

\bibitem[Bank et~al., 2016]{voss2016}
Bank, P., Soner, H.~M., \& Vo{\ss}, M. (2016).
\newblock Hedging with temporary price impact.
\newblock {\em Mathematics and Financial Economics}.
\newblock to appear.

\bibitem[Chan \& Sircar, 2015]{chan2015optimal}
Chan, P. \& Sircar, R. (2015).
\newblock Optimal trading with predictable return and stochastic volatility.

\bibitem[Crandall \& Lions, 1987]{crandall1987}
Crandall, M.~G. \& Lions, P.-L. (1987).
\newblock Remarks on the existence and uniqueness of unbounded viscosity
  solutions of hamilton-jacobi equations.
\newblock {\em Illinois J. Math.}, 31(4):665--688.

\bibitem[Davini \& Kosygina, 2017]{kosygina2017}
Davini, A. \& Kosygina, E. (2017).
\newblock Homogenization of viscous and non-viscous {HJ} equations: a remark
  and an application.
\newblock {\em Calculus of Variations and Partial Differential Equations}, 56.

\bibitem[Davis \& Norman, 1990]{davisNorman1990}
Davis, M. H.~A. \& Norman, A.~R. (1990).
\newblock Portfolio selection with transaction costs.
\newblock {\em Mathematics of Operations Research}, 15(4):676--713.

\bibitem[Fleming \& Soner, 2006]{flemingSoner2006}
Fleming, W. \& Soner, H. (2006).
\newblock {\em Controlled Markov Processes and Viscosity Solutions}.
\newblock Applications of mathematics. Springer.

\bibitem[Fouque et~al., 2001]{Fouque2001}
Fouque, J.-P., Papanicolaou, G., \& Sircar, K.~R. (2001).
\newblock From the implied volatility skew to a robust correction to
  {B}lack-{S}choles {A}merican option prices.
\newblock {\em International Journal of Theoretical and Applied Finance},
  4(4):651--675.

\bibitem[Fouque et~al., 2011]{fouque2011multiscale}
Fouque, J.-P., Papanicolaou, G., Sircar, R., \& S{\o}lna, K. (2011).
\newblock {\em Multiscale Stochastic Volatility for Equity, Interest Rate, and
  Credit Derivatives}.
\newblock Cambridge University Press.

\bibitem[G{\^a}rleanu \& Pedersen, 2013]{garleanuPedersen}
G{\^a}rleanu, N. \& Pedersen, L.~H. (2013).
\newblock Dynamic trading with predictable returns and transaction costs.
\newblock {\em The Journal of Finance}, 68(6):2309--2340.

\bibitem[Grossman \& Zhou, 1993]{grossmanZhou}
Grossman, S.J.  \& Zhou, Z. (1993).
\newblock Optimal investment strategies for controlling drawdowns.
\newblock {\em Mathematical Finance}, 3(3):241--276.


\bibitem[Howison, 2005]{howison2005practical}
Howison, S. (2005).
\newblock {\em Practical Applied Mathematics: Modelling, Analysis,
  Approximation}.
\newblock Cambridge Texts in Applied Mathematics. Cambridge University Press.


\bibitem[Jarrow et~al., 2004]{article}
Jarrow, R., Cetin, U., \& Protter, P. (2004).
\newblock Liquidity risk and arbitrage pricing theory.
\newblock 8:311--341.

\bibitem[Kabanov \& Safarian, 2008]{kabanovMarkets2008}
Kabanov, Y. \& Safarian, M. (2008).
\newblock {\em Markets with Transaction Costs}.
\newblock Springer, Berlin, London.

\bibitem[Obizhaeva \& Wang, 2013]{OW2013}
Obizhaeva, A. A. \& Wang, J. (2013). 
\newblock Optimal trading strategy and supply/demand dynamics.
\newblock {\em Journal of Financial Markets}, 16(1):1--32.

\bibitem[Rogers \& Singh, 2010]{rogers2010}
Rogers, L. C.~G. \& Singh, S. (2010).
\newblock The cost of illiquidity and its effects on hedging.
\newblock {\em Mathematical Finance}, 20(4):597--615.

\bibitem[Shreve \& Soner, 1994]{shreve1994}
Shreve, S.~E. \& Soner, H.~M. (1994).
\newblock Optimal investment and consumption with transaction costs.
\newblock {\em Ann. Appl. Probab.}, 4(3):609--692.

\end{thebibliography}
\end{document}